\RequirePackage{fix-cm}
\documentclass[smallextended,twocolumn]{revtex4}       
\usepackage{bbm}
\usepackage{graphicx}
\usepackage{mathptmx}
\usepackage{amssymb,amsmath,amsfonts,latexsym}
\def \beq {\begin{equation}}
\def \eeq {\end{equation}}
\def \tr {\rm Tr}
\newcommand {\qs} {{\rm Q}_{\rm S}}
\newcommand {\qt} {{\rm Q}_{\rm T}}
\newcommand {\ks} {k_{\rm S}}
\newcommand {\kt} {k_{\rm T}}
\usepackage{graphicx}

\begin{document}

\title{Revealing the properties of the radical-pair magnetoreceptor using pulsed photo-excitation timed with pulsed rf}


\author{K. Mouloudakis and I. K. Kominis}
\affiliation{ Department of Physics, University of Crete, Heraklion 71003, Greece \\\email{ikominis@physics.uoc.gr}}


\begin{abstract}
The radical-pair mechanism is understood to underlie the magnetic navigation capability of birds and possibly other species. Experiments with birds have provided indirect and in cases conflicting evidence on the actual existence of this mechanism. We here propose a new experiment that can unambiguously identify the presence of the radical-pair magnetoreceptor in birds and unravel some of its basic properties. The proposed experiment is based on modulated light excitation with a pulsed laser, combined with delayed radio-frequency magnetic field pulses.  We predict a resonance effect in the birds' magnetic orientation versus the rf-pulse delay time. The resonance's position reflects the singlet-triplet mixing time of the magnetoreceptor. 
\end{abstract}
\maketitle
\section{Introduction}
Animal magnetoreception \cite{quinn,phillips,johnsen,qin} and specifically avian magnetoreception \cite{WW,WWscience,mouritsenR,horePNAS,ritz2000} is a long-standing and still unresolved scientific puzzle. A wealth of data \cite{green,ww78,munro,borland,sandberg,wild,mour2009,mour2010,mour2012} has made the magnetic navigation capabilities of birds unquestionable. However, the particular mechanism underlying this capability remains elusive. Magnetite crystals in the bird's upper beak \cite{m1,m2,m3,m4,m5} and the photo-initiated radical-pair mechanism \cite{schulten1978} in the avian retina are the two prevalent hypotheses behind the biophysical realization of avian magnetoreceptors. Regarding the latter, the specific radical-pair (RP) magnetoreceptor is still unknown, even though cryptochrome has been a major protein candidate supporting magnetic sensitive RP reactions \cite{cry1,cry2,cry3,cry4}. 

A significant experimental signature of the RP mechanism was the radio-frequency resonance effect \cite{ritz_nature}, where radio-frequency (rf) magnetic fields transverse to the static field and of particular frequencies were shown to disorient the birds. This directly pointed to the RP mechanism since the molecule-specific electron spin resonances are expected to be excited by resonant rf fields. However, a recent experiment studying rf disorientation could not reproduce this resonance effect \cite{mouritsen_2014}. Moreover, the magnitude of the disorienting rf fields used in \cite{ritz_nature,mouritsen_2014} is far smaller than theoretically required by the RP mechanism \cite{kavokin}. To our understanding, experiments with cw light excitation and cw magnetic noise have reached their limits in how much more information they can extract. It thus appears that further progress in making a convincing case for the RP compass requires new experimental signatures. 

We here propose a new experiment using pulsed photo-excitation combined with pulsed rf magnetic fields, in a way that can unambiguously identify the presence of the radical-pair compass and extract its basic parameters. In Section 2 we discuss the RP model used for the analysis. In Section 3 we proceed to examine pulsed photoexcitation pulses followed by pulsed rf magnetic fields, the rf pulses following the laser pulses by a variable delay time. Singlet RPs are insensitive to magnetic fields, while triplet RPs are randomized by the rf magnetic fields. Hence only when the rf pulse is delayed with respect to the laser pulse by the S-T mixing time will one observe the disorientation of the compass. In Section 4 we discuss the experimental implementation.

\section{Radical-pair model used for the simulations}
\begin{figure}
\begin{center}
\includegraphics[width=7 cm]{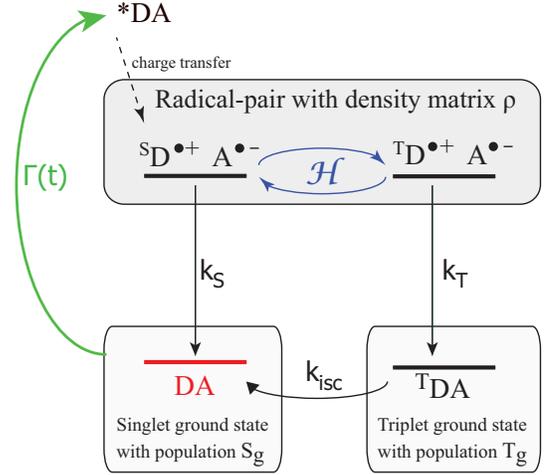}
\caption{Radical-pair reaction dynamics. The population of the singlet donor-acceptor precursor DA is considered to be the signal carrying the magnetic field information into deeper stages of neural processing. This population is drained by  photoexcitation at the rate $\Gamma(t)$, which in this particular work is time-dependent. It is increased by the radical-pair singlet recombination and by the intersystem crossing from the triplet ground state, introduced in order to close the reaction. The singlet and triplet recombination rates are $\ks$ and $\kt$, respectively, and ${\cal H}$ is the magnetic Hamiltonian inducing singlet-triplet oscillations between the singlet and triplet radical-pairs,  $^{\rm S}{\rm D}^{\bullet +}{\rm A}^{\bullet -}$ and  $^{\rm T}{\rm D}^{\bullet +}{\rm A}^{\bullet -}$. The charge transfer from the photo-excited molecule $^{*}$DA is much faster than all other rates, hence the rate of creation of radical-pairs is effectively $\Gamma$.}
\label{model}
\end{center}
\end{figure}

We use a simple RP model to produce the simulations conveying the idea behind the proposed experiment. In particular, we consider an RP with one nuclear spin in the donor molecule, having an anisotropic hyperfine coupling with the donor's electron. The hyperfine tensor is considered to have
$A_{xx}=A$ and all other elements zero, thus the magnetic Hamiltonian is
\beq
{\cal H}=\omega\big(\cos\phi(s_{1x}+s_{2x})+\sin\phi(s_{1y}+s_{2y})\big)+As_{1x}I_{x}\label{Hm}
\eeq
Here $\omega$ is the electron Larmor frequency in the applied static magnetic field, taken to be on the x-y plane, $s_{1i}$ and $s_{2j}$ refer to the $i$-th and $j$-th spin component of the donor's and acceptor's electron, respectively, and $I_x$ is the x-component of the donor's single nuclear spin. The other pertinent rates are seen in Fig.\ref{model}. The singlet and triplet recombination rates are taken equal and denoted by $k$. To close the reaction we also consider an intersystem crossing rate $k_{\rm isc}$ transforming triplet neutral products into the singlet precursors. Light excites the ground state DA molecules to $^{*}$DA at a rate $\Gamma$, and charge transfer leads to the creation of singlet RPs. Since the rate of the latter process is \cite{cry4} much larger than $\Gamma$ and all other rates of the problem, the rate of RP creation is $\Gamma$. For the same reason, i.e. the fact that the population of $^{*}$DA is drained practically instantaneously, there is no need to consider stimulated emission of the exciting light.

The population of the singlet precursors DA is taken to be the signaling state carrying the magnetic field information towards further neural processing leading to the bird's orientation. 
In many spin-chemistry calculations the RPs are considered to be all initialized in the singlet state at time $t=0$ and one then calculates the reaction yields resulting at the end of a single reaction cycle. For this work, however, we need to continuously create RPs at a rate $\Gamma$ and calculate the steady-state population of the neutral DA molecules, $S_{\rm g}$, in the scheme of a continuously running and closed reaction of Fig.\ref{model}. To do so, we add a source term to the Haberkorn master equation for the RP density matrix $\rho$:
\beq
{{d\rho}\over {dt}}=\Gamma S_{\rm g}\rho_0-i[{\cal H},\rho]+{\cal R}(\rho),\label{drdt}
\eeq
where $\rho_0=\qs/\tr\{\qs\}$ is the initial density matrix of singlet RPs having zero nuclear spin polarization, and
\beq
{\cal R}(\rho)=-{\ks\over 2}(\qs\rho+\rho\qs)-{\kt\over 2}(\qt\rho+\rho\qt)
\eeq
is the reaction super-operator describing singlet and triplet RP recombination. We used the traditional (Haberkorn) master equation, since any quantum effects \cite{review} beyond this approach are not relevant to this work. Nevertheless, we checked the results of our master equation, involving singlet-triplet decoherence, and they are qualitatively the same. 
The first term in Eq. \eqref{drdt} creates $\Gamma S_{\rm g}$ RPs per unit time in the state $\rho_0$. To close the reaction we also 
consider the following two rate equations for $S_{\rm g}$ and the corresponding triplet ground state population, $T_{\rm g}$:
\begin{align}
{{dS_{\rm g}}\over {dt}}&=-\Gamma S_{\rm g}+\ks\tr\{\qs\rho\}+k_{isc}T_{\rm g}\label{dsgdt}\\
{{dT_{\rm g}}\over {dt}}&=\kt\tr\{\qt\rho\}-k_{isc}T_{\rm g}\label{dtgdt}
\end{align}
The first of the above equations describes the depopulation of $S_{\rm g}$ by photoexcitation at the rate $\Gamma$ and the population of $S_{\rm g}$ by (i) the singlet RP recombination and (ii) the intersystem crossing from $^{\rm T}$DA at the rate $k_{\rm isc}$. The second describes the depopulation of $^{\rm T}$DA at the rate $k_{\rm isc}$ and its population by the triplet RP recombination. Finally, when solving the system of equations \eqref{drdt}, \eqref{dsgdt} and \eqref{dtgdt}, the initial condition is $S_{\rm g}(t=0)=1$.

Before moving to the main part of this work, i.e. the pulsed photoexcitation for which the excitation rate $\Gamma$ is time-dependent, we first discuss the continuous illumination case $\Gamma={\rm const}$ in order to get some insight into the quantities of interest. We first note that in our numerical work (except for the Hamiltonian evolution of Fig.\ref{STmixing}) all rates will be given relative to the recombination rate $k=\ks=\kt=1$. Accordingly, time will have units $1/k=1$.
 
In Fig. 2 we plot the steady-state population $S_{\rm g}$, evaluated numerically from  \eqref{drdt}, \eqref{dsgdt} and \eqref{dtgdt}, as a function of $\phi$ for two values of constant $\Gamma$, where $\phi$ is the angle between the magnetic field (lying on the x-y plance) and the x-axis defining the hyperfine anisotropy. The avian compass is based on the $\phi$-modulation of the population $S_{\rm g}$. We define 
\beq
\Delta S\equiv{{{\rm max}_{\phi}\{S_{\rm g}\}-{\rm min}_{\phi}\{S_{\rm g}\}}\over {{\rm max}_{\phi}\{S_{\rm g}\}+{\rm min}_{\phi}\{S_{\rm g}\}}}
\eeq
and call it $\phi$-visibility. The measured heading error in experiments with birds is inversely proportional to $\Delta S$. It is seen that the higher $\Gamma$, the faster is drained the ground state DA, hence the smaller its steady state population. For the pulsed photo-excitation we use an average excitation rate $\overline{\Gamma}=0.25$.
\begin{figure}
\begin{center}
\includegraphics[width=8 cm]{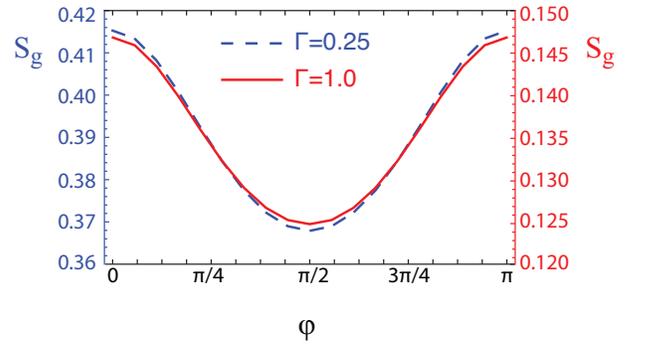}
\caption{Angular modulation of the singlet ground state (DA) population for two different values of a constant excitation rate, $\Gamma=0.25$ (dashed blue line) and $\Gamma=1.0$ (solid red line). The parameters of the RP model are $\omega=\ks=\kt=1$, $A=10$ and $k_{\rm isc}=0.1$. For the higher excitation rate the state DA is depleted faster and hence both the population $S_{\rm g}$ and the difference ${\rm max}_{\phi}\{S_{\rm g}\}-{\rm min}_{\phi}\{S_{\rm g}\}$ become smaller. }
\label{cont}
\end{center}
\end{figure}
What is of interest for the time-delay resonance effect to be presented in the following is the time evolution of the RP state resulting just from the Hamiltonian term in the master equation \eqref{drdt}. Using this Hamiltonian time evolution, we plot in Fig.\ref{STmixing} the triplet state probability $\langle\qt\rangle$ as a function of time for three different angles $\phi$. It is seen that the first instance in time when the triplet state is reached, i.e. when $\langle\qt\rangle\approx 1$, is largely independent of $\phi$ and, as expected, scales as $1/A$.
\begin{figure}
\begin{center}
\includegraphics[width=8 cm]{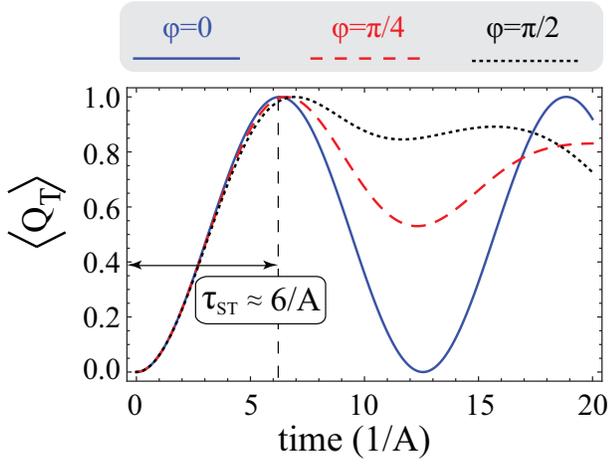}
\caption{Singlet-triplet mixing driven by the Hamiltonian ${\cal H}$ of Eq. \eqref{Hm} with $\omega=1$. We plot the triplet expectation value $\langle\qt\rangle$ as a function of time (in units of $1/A$) for three different angles $\phi$. It is seen that the first instance of S-T conversion is independent of $\phi$ and takes place at a time $\tau_{\rm ST}\approx 6/A$ for this particular Hamiltonian.}
\label{STmixing}
\end{center}
\end{figure}
\section{Photoexcitation pulses followed by rf pulses}
\begin{figure}
\begin{center}
\includegraphics[width=8 cm]{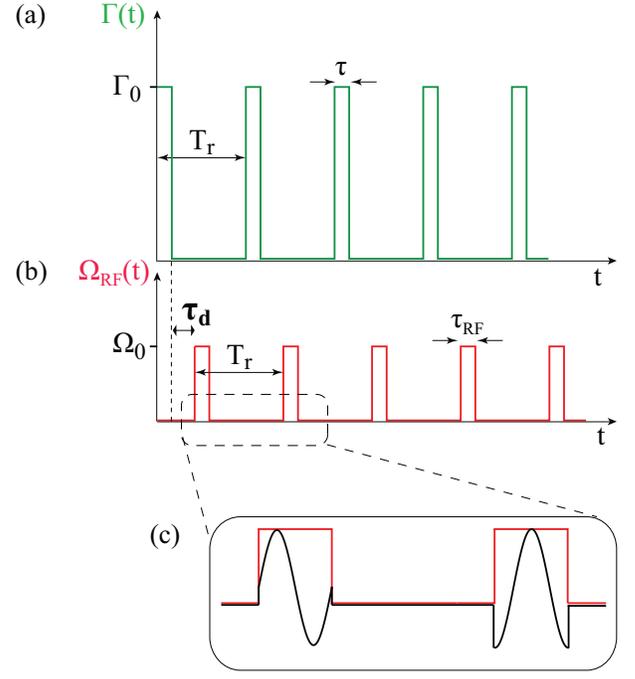}
\caption{(a) Photo-excitation rate $\Gamma(t)$, consisting of a pulse train with pulse amplitude $\Gamma_0$, pulse width $\tau$ and pulse repetition time $T_r$. (b) Envelope of the rf field $\Omega_{\rm rf}(t)$, consisting of a pulse train with pulse amplitude $\Omega_0$, pulse width $\tau_{\rm rf}$ and pulse repetition time $T_r$. This pulse train is delayed from the photo-excitation pulse train by $\tau_d$. (c) Rf carrier wave modulated by the envelope shown in (b). In order for the rf frequency spectrum to be continuous and simulate noise we insert a random pulse-to-pulse phase difference $\psi$. }
\label{exc}
\end{center}
\end{figure}

We will here provide a detailed analysis of  the idea of the proposed experiment. There are three main ingredients to this idea. First, as well known, the singlet state is not sensitive to any magnetic field, constant or alternating. The mechanism through which the avian RP compass is disoriented by rf fields necessarily starts with the induced spin randomization of the triplet state. Second, if the photo-excitation is pulsed, the transformation of singlet RPs to triplet RPs takes place in well defined times, given the S-T mixing frequency $\Omega_{\rm ST}$. Third, if the radio frequency pulses are delayed with respect to the light pulses, as shown in Fig.\ref{exc}, it is expected that by varying the delay time $\tau_d$, the birds' magnetic orientation, as measured by $\Delta S$, will exhibit a resonance, as an increasing delay will correspond to an increasingly triplet character of the RP's spin state. The resonance dip will happen at a particular delay $\tau_d$ such that the RPs that were photo-excited to the singlet state will have oscillated into a predominantly triplet spin character. Observing this resonance dip will thus (i) unambiguously reveal the presence of the radical-pair magnetoreception mechanism and (ii) unravel the mixing frequency $\Omega_{\rm ST}$ of the particular magneto receptor molecule. 

The above picture is exemplified in the following. The photo-excitation rate $\Gamma(t)$ is shown in Fig.\ref{exc}a. It consists of pulses of amplitude $\Gamma_0$, pulse width $\tau$ and repetition time $T_r$. The amplitude of the photo-excitation pulses, $\Gamma_0$, is given a value such that the time average $\overline{\Gamma}$ of $\Gamma(t)$ is the same as the $\Gamma=0.25$ case of continuous excitation shown in Fig.\ref{cont}. We choose $\tau=0.005$ for the pulse width and $T_r=2$ for the pulse repetition time, hence $\Gamma_0=\overline{\Gamma}T_r/\tau=100$. 

To include the presence of the pulsed rf we add to the Hamiltonian \eqref{Hm} the term
\beq
{\cal H}_{\rm rf}=\Omega_{\rm rf}(t)\cos(\omega_{\rm rf}t+\psi)\big(s_{1z}+s_{2z}\big)
\eeq

We took the rf magnetic field to be polarized along the z-axis, perpendicular to the static magnetic field lying on the x-y plane. $\Omega_{\rm rf}(t)$ is the pulse train envelope shown in Fig.\ref{exc}b. The pulse amplitude and width are $\Omega_0$ and $\tau_{\rm rf}$, respectively. The pulse delay time with respect to the photoexciation pulses is $\tau_d$, which is variable. The pulse repetition time is the same as for $\Gamma(t)$, i.e. $T_r$. 
The amplitude of the rf magnetic field, given in terms of its Rabi frequency $\Omega_0$, is taken $\Omega_0=15\omega$, i.e. the rf-field amplitude is 15 times earth's field. We note that this is way higher than the rf-field amplitudes experimentally found to disorient the birds. As mentioned in the introduction and clearly stated in \cite{mouritsen_2014}, it is still an unresolved puzzle why the {\it theoretically} required rf-field amplitude is so much higher than what is experimentally observed to disorient the birds. We further elaborate on this point in the following Section on the experimental implementation. Finally, we take $\tau_{\rm rf}=0.1$. 

The rf carrier we use, shown in Fig.\ref{exc}c, is a cosine wave of frequency $\omega_{\rm rf}=20$. In the experiment one must use pulsed noise of a bandwidth similar to \cite{mouritsen_2014}. To simulate that theoretically we include a pulse-to-pulse random phase $\psi$ in the cosine wave. Without this phase the rf pulse train would have a discrete Fourier spectrum. With the inclusion of these random phases we theoretically simulate the pulsed rf noise since now the Fourier spectrum of $\Omega_{\rm rf}(t)\cos(\omega_{\rm rf}t+\psi)$ is continuous and has a bandwidth of about $1/\tau_{\rm rf}$. 
\begin{figure}
\begin{center}
\includegraphics[width=8 cm]{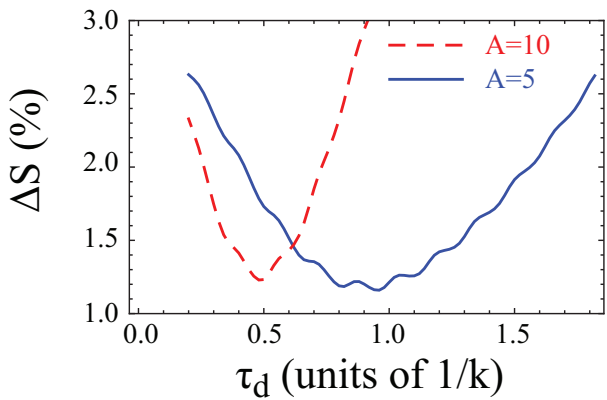}
\caption{Time-delay resonance effect predicted in this work. Shown is the $\phi$-visibility as a function of the time delay $\tau_d$ of the rf-pulses with respect to the laser pulses, for two different values of the hyperfine coupling $A$. For the radical-pair we took $k\equiv\ks=\kt=1$, $k_{\rm sic}=0.1$ and $\omega=1$. For the pulse trains we took $\Gamma_0=100$, $\tau=0.005$, $T_r=2$, $\tau_{\rm rf}=0.1$, $\Omega_{\rm rf}=15$ and $\omega_{\rm rf}=20$. It is seen that for higher $A$, singlet-triplet mixing is taking place faster, hence the time-delay required to hit the triplet state is smaller. For zero time delay the $\phi$-visibility for this model is about 3\%, and at the resonance dip it falls by a factor of 3 for the chosen value of $\Omega_{\rm rf}$. }
\label{res}
\end{center}
\end{figure}
In Fig.\ref{res} we depict the time-delay resonance effect. The change of $\Delta S$ from the off-resonant to the on-resonant time delay is significant enough (about a factor of 3) that the compass should disorient on resonance. We see that by varying the hyperfine coupling $A$ the resonance's position is shifted in accordance with Fig.\ref{STmixing}. That is, according to Fig.\ref{STmixing}, the S-T mixing time is about $6/A$, and for the two values used for the hyperfine coupling, $A=5$ and $A=10$, the position of the time-delay resonance is $\tau_d\approx 1$ and $\tau_d\approx 0.5$, respectively. The different resonance width observed in Fig.\ref{res} is due to the different interplay of the S-T mixing (dependent on $A$) with the pulse repetition time $T_r$. We finally note that Fig.\ref{res} was produced by a moving average of the actual result in order to remove a (still visible) modulation artifact stemming from the numerical scanning of the delay time $\tau_d$. 

We have checked that the resonance phenomenon persists for a multi-nuclear spin radical pair. In particular, we run the same simulation for a radical-pair containing up to 4 nuclear spins. We note that by choosing the relevant hyperfine couplings so that the angular modulation of Fig. 2 is significant, we also obtain a significant resonance dip like in Fig. 5. In other words, it appears that if the compass has evolved to reach an optimum angular yield dependence, it will exhibit the resonance effect we presented.  
On the other hand, by no means do we claim that the effect will be experimentally detected no matter what. What we claim is that this is a viable measurement to do with live birds, and if the resonance phenomenon is realized, it will provide for a clean and information-rich signature of the radical-pair magnetoreceptor.
\section{Experimental Implementation}
All rates of the problem have been expressed relative to the recombination rate $k$, which was given the value 1. For the following numerical estimates we take $1/k=1~\mu{\rm s}$. In any case, an educated guess of $k$ must be made in order to set the timescale of the experiment.
\subsection{Laser Pulses} 
Pulsed lasers with pulse duration on the order of 1-10 ns, a repetition rate on the order of 200-500 kHz and a wavelength within the sensitivity window of the avian magnetoreceptor are commercially available.
The pulsed laser can be fed into a diffuser and illuminate the birds' cage just like the regular illumination with lamps or diodes. For ns lasers, any pulse broadening by the diffuser is negligible given the much slower reaction and magnetic dynamics. In other words, since we took $\tau=0.005$ (in units of $1/k$) for the laser pulse width, any pulse broadening will leave the pulse width still much smaller than the magnetic and recombination dynamics taking place at the timescale $1/k=1$. Regarding the laser pulse peak intensity, in the case of continuous illumination a flux of about $10^{16}$ photons/s/${\rm m}^2$ is known \cite{wiltschko_I1,wiltschko_I2} to be enough for the compass to function. Assuming a total illumination area on the order of 1 ${\rm m}^2$, the light source's average power should then be about 5 mW (at 500 nm). We took the pulse width to be 400 times smaller than the pulse repetition time, so to get the same average photoexcitation rate the pulse peak power should be 2 W. For a 1 ns pulse this translates into a pulse energy of 2 nJ, which is well within the capabilities of commercially available and simple table-top lasers. 
\subsection{Radio-frequency pulses}
In our calculations we took the rf pulse width to be $\tau_{\rm rf}=0.1$, which is small enough compared to a typical mixing frequency $\Omega_{\rm ST}\approx 1$ (see Fig.\ref{STmixing}). This pulse width translates to 100 ns. In producing Fig.\ref{res} we scanned the delay time in steps of 0.02, translating to 20 ns. To summarize, we need 50-100 ns wide rf pulses modulating noise of bandwidth of  about 10 MHz, the delay of the pulses being scanned in steps of about 20-50 ns. Such rf pulse generators are commercially available. Similarly, the power of the rf magnetic field should be the one used in \cite{mouritsen_2014} scaled up by the ratio $T_r/\tau_{\rm rf}\approx 20$ since now we have pulsed and not continuous rf. Again, this is readily achievable. 
\section{Conclusions}
We have proposed an experiment using pulsed photo excitation in conjunction with properly delayed pulses of radio frequency magnetic fields to study the response of avian magnetic orientation. If the radical-pair mechanism is indeed responsible for the avian compass, a robust resonance will appear in the measured birds' orientation versus delay time between laser and rf pulses. Further, the particular delay time at the resonance's dip is the inverse of the singlet-triplet mixing frequency of the magneto receptor molecule. We analyzed this experiment using a generic radical-pair model, but the realization of the experiment as well as the result we obtained for the time-delay resonance effect is robust and independent of the particular radical-pair model. For example, one could consider an RP with just one non-zero recombination rate, e.g. the singlet, and no intersystem-crossing. The singlet ground state population would again be the signaling state, depending on $\phi$ through the different time spent by the RP in the triplet state. Similar results would be obtained in this case. The same experiment could also be used for other magneto receptive species \cite{flies,flies2,mouse} in which the RP mechanism is presumed to exist.
\begin{acknowledgements}
We acknowledge support from the European Union's Seventh Framework Program FP7-REGPOT-2012-2013-1 under grant agreement 316165.
\end{acknowledgements}


\begin{thebibliography}{}

\bibitem{quinn}
T. P. Quinn and E. L. Brannon, J. Comp. Physiol. {\bf 147}, 547 (1982).

\bibitem{phillips}
J. B. Phillips and S. C. Borland, Nature {\bf 359}, 143 (1992).

\bibitem{johnsen}
S. Johnsen and K. J. Lohmann, Phys. Today {\bf 61} (3), 29 (2008).

\bibitem{qin}
S. Qin {\it et al.}, Nature Mat. {\bf 15}, 217 (2016). 

\bibitem{WW}
R. Wiltschko and W. Wiltschko, Animal Behaviour {\bf 65}, 257 (2003). 

\bibitem{WWscience}
W. Wiltschko and R. Wiltschko, Science {\bf 176}, 62 (1972).

\bibitem{mouritsenR}
H. Mouritsen, Nature {\bf 484}, 320 (2012).

\bibitem{horePNAS}
D. T. Rodgers and P. J. Hore, Proc. Natl. Acad. Sci. USA {\bf 106}, 353 (2009).

\bibitem{ritz2000}
T. Ritz, S. Adem and K. Schulten, Biophys. J. {\bf 78}, 707 (2000).

\bibitem{green}
C. Walcott and R. P. Green, Science {\bf 184}, 180 (1974).

\bibitem{ww78}
R. Wiltschko and W. Wiltschko, Naturwissenschaften {\bf 65}, 112 (1978).

\bibitem{munro}
U. Munro, J. A. Munro, J. B. Phillips, R. Wiltschko and W. Wiltschko, Naturwissenschaften {\bf 84}, 26 (1997).

\bibitem{borland}
M. E. Deutschlander, J. B. Phillips and S. C. Borland, J. Exp. Biol. {\bf 202}, 891 (1999).

\bibitem{sandberg}
G. A. Gudmundsson and R. Sandberg, J. Exp. Biol. {\bf 203}, 3137 (2000).

\bibitem{wild}
M. N. Williams and J. M. Wild, Brain Res. {\bf 889}, 243 (2001).

\bibitem{mour2009}
M. Zapka {\it et al.}, Nature {\bf 461}, 1274 (2009).

\bibitem{mour2010}
D. Heyers, M. Zapka, M. Hoffmeister, J. M. Wild and H. Mouritsen, Proc. Natl. Acad. Sci. USA {\bf 107}, 9394 (2010).

\bibitem{mour2012}
H. Mouritsen and P. J. Hore, Current Opinion in Neurobiology {\bf 22}, 343 (2012).

\bibitem{m1}
J. L. Kirschvink and J. L. Gould, Biosystems {\bf 13}, 181 (1981).

\bibitem{m2}
V. P. Shcherbakov and M. Winklhofer, Eur. Biophys. J. {\bf 28}, 380 (1999).

\bibitem{m3}
A. F. Davila, M. Winklhofer, V. P. Shcherbakov and N. Peterson, Biophys. J. {\bf 89}, 56 (2005).
. 
\bibitem{m4}
I. A. Solov¢yov and W. Greiner, Biophys. J. {\bf 93}, 1493 (2007).

\bibitem{m5}
M. M. Walker, J. Theor. Biol. {\bf 250}, 852008 (1999).

\bibitem{schulten1978}
K. Schulten, C. E. Swenberg and A. Weller, Z. Phys. Chem. {\bf 111}, 1 (1978).

\bibitem{cry1}
A. Cashmore, J. Jarillo, Y. J. Wu and D. Liu, Science {\bf 284}, 760 (1999).

\bibitem{cry2}
H. Mouritsen {\it et al.},  Proc. Natl. Acad. Sci. USA {\bf 101}, 14294 (2004).

\bibitem{cry3}
I. A. Solov¢yov, D. Chandler and K. Schulten, Biophys. J. {\bf 92}, 2711 (2007).

\bibitem{cry4}
I. A. Solov¢yov and K. Schulten, J. Phys. Chem. B {\bf 116}, 1089 (2012).

\bibitem{ritz_nature}
T. Ritz, P. Thalau, J. B. Phillips, R. Wiltschko and W. Wiltschko, Nature {\bf 429}, 177 (2004).

\bibitem{mouritsen_2014}
S. Engels {\it et al.}, Nature {\bf 509}, 353 (2014).

\bibitem{kavokin}
K. V. Kavokin, Bioelectromagnetics {\bf 30}, 402 (2009).

\bibitem{review} 
I. K. Kominis, Mod. Phys. Lett. B {\bf 29}, 1530013 (2015).

\bibitem{WWintensity}
R. Wiltschko, K. Stapput, H.-J. Bischof and W. Wiltschko, Frontiers in Zoology {\bf 4} (2007) 5.

\bibitem{wiltschko_I1}
W. Wiltschko, R. Wiltschko and U. Munro, Naturwissenschaften {\bf 87}, 36 (2000).

\bibitem{wiltschko_I2}
W. Wiltschko and R. Wiltschko, Naturwissenschaften {\bf 89}, 445 (2002).

\bibitem{flies}
R. J. Gegear, L. E. Foley, A. Casselman and S. M. Reppert, Nature {\bf 463}, 804 (2010).

\bibitem{flies2}
B. Paulus {\it et al.}, Febs {\bf 282}, 3175 (2015).

\bibitem{mouse}
E. P. Malkemper {\it et al.}, Sci. Rep. {\bf 4}, 9917 (2015).

\end{thebibliography}
\end{document}